# Exploring Water Governing System Fit Through a Statistical Mechanics Approach

Peyman Arjomandi A.[a,†,*], Seyedalireza Seyedi[b,c,†], Ehsan Tavakoli Nabavi[d, ††], Saeid Alikhani[b, ††]

[a]*Department of Civil, Chemical, Environmental and Materials Engineering (DICAM), University of Bologna, Bologna, Italy*
[b]*Department of Mathematical Sciences, Yazd University, Yazd, Iran*
[c]*Mathematical Modelling Centre, Yazd University, Yazd, Iran*
[d]*National Centre for the Public Awareness of Science, Australian National University, Canberra, Australia.*

**Abstract**

Water governing systems are twisted with complex interplays among levels and scales which embody their structures. Typically, the mismatch between human-generated and natural systems produces externalities and inefficiencies reflectable in spatial scales. The largely known problem of fit in water governance is investigated to detect the issues of fit between administrative/institutional scales and the hydrological one in a lake basin. To implement the idea, constraining the level of analysis interlinked to the concentrated levels of administration in spatial scales, the fit of the governing system was analyzed by means of statistical mechanics. Modeling the structure of water demand/supply governing system in a given region through the Curie-Weiss Mean Field approximation, the system cost in relation to its structure and fit was appraised and compared with two other conceptual structures in the Urmia Lake Basin in Iran. The methodology articulated an analysis framework for exploring the effectiveness of the formulated water demand/supply governing system and its fit to the relevant hydrological system. The findings of this study may help developing strategies to encourage adaptations, rescaling/reforms for effective watershed management.

## 1. Introduction:

Management of water resources and water governance confront with the challenge of inherent scales and levels playing an intricate role in their scope of steer. Due to the diversity of parameters affecting the water governing systems, the hydrological knowhow solely is not able to satisfy their proper formulation, and to increase the effectiveness of their mechanism the incorporation of knowledge of different disciplines are imperative. To address the complexity in such systems there is a need to pinpoint the scale and cross-scale dynamics (Cash et al., 2006).

* Corresponding author:
*E-mail address:* peyman.arjomandi2@unibo.it (P. Arjomandi A.)
† These authors contributed equally to this work as the first author
†† These authors contributed equally to this work as the second author

Some authors (e.g. Gibson et al., 2000; Cash et al., 2006) have defined these scales as the spatial, temporal, jurisdictional, institutional, and other dimensions which are used to study or measure relevant phenomenon in their domain. Besides, levels communicate the units of analysis within the scope of scales. Also, authors have used the term scale when a graduated range of extent is dealt with, and levels are notified as non-continuous classes in the realm of scales (Daniell and Barreteau, 2014). Although in categorial scope, scales are independent, they are combinable with each other. This aspect is well noted regarding the spatial and administrative scales where conventionally can be found both together. Such endowment is usual because various spatial levels associated to jurisdictions respective to their administration levels.

According to Cash et al. (2006), a society may tackle with three common challenges within the scale's arena. The first one stems of scales/levels interactions which may fail to be recognized. The second one arises from mismatches between the scales of human-adjusted and natural systems, and the third, is generated by the heterogeneous values that disparate actors may assign to different scales/levels. Indeed, all the challenges are awkward. However, the second challenge is debated by many authors in socio-ecological, and water governance subjects which can result in externalities and inefficiencies (e.g. Young, 2002; Folke et al., 2007; Moss and Newig, 2010).

Water as a natural element with flowable characteristics in spatial scales has been under tenure tensions since millenniums. The physical flow of water has a tight relationship with political, economic, social and human flows (for more info please see Daniell and Barreteau, 2014). Therefore, hydrological (spatial) scales such as catchments, basins and so on, commonly face governance conflicts. While the recent developments in human civilization and diplomacy have tried to deploy appropriate solutions for short as well as long term agreements on utilization of water resources, due to pollutions, depletions and other issues of water resources, besides the climate change impacts, still there are obstacles and difficulties to place suitable and sustainable cooperation mechanisms for extraction and use of water (Zareie et al., 2020).

Tackling with competing scales, typically the structure of water governance is determined by the political, economic, and administrative systems (Rogers and Hall, 2003). Majorly, in this area, water demand-supply and allocation plans are considered based on the need of users rather than the natural availability of water resource. Therefore, the system of governance is influenced by decisions at administrative/institutional scales and at the same time affects the flows of water and watershed scales. This issue generally creates conflicts due to externalities between the natural governing system (biophysical/hydrological) and human-tailored one (Young, 2006). Since, conventionally, decisions are substantially oriented by the dominancy of human-determined objectives in comparison to environmental requirements, it leads to unfit of such administrative-political systems to hydrological basins.

The question may arise here is that how to distinguish the more match governance system to a given hydrological system. In other words, how to figure out the more fit administrative/institutional system to a watershed. Answering such question is controversial and deals with large complexity. It needs involvement of the opinions and achievements of many disciplines and deals with various dimensions. However, some arrangements can help observing the externalities or mismatches in focused levels of systems. In line with that, constraining the levels of analysis, a particular lens should be adopted suitable to follow the relevant flows in certain levels of contributing scales (Gibson et al., 2000).

A governing system is recognized by its jurisdictional boundaries implicit in spatial scales. The levels associated to such scales could be for instance local, provincial, national etc., and on the other hand, such levels wholly, partially or jointly meet up with levels of hydrological scales such as catchment, river basin, lake basin etc. which nevertheless are types of spatial scales. Therefore, both human-generated and biophysical governing systems launch relevant levels on spatial scales respective to their pertinent boundaries. For instance, a regional/provincial level of a political-administrative scale which spatially has been furnished surrounding a lake basin (a level of hydrological scale), could be detected through its jurisdictional boundary alongside a particular spatial level. Such boundary in that level is typically the borders of a politically/institutionally partitioned region (e.g. provincial borders). Such district may include several geographical zones such as urban or rural areas trying to obtain their water requirements through the routes created by administrative/institutional systems. A given task/service there, like as water demand/supply is operated through a mechanism in which the routes for requesting and delivering a service reveal the (administrative) structure for flows in associated levels. Hence, monitoring such routes the effectiveness of the devised administrative levels for water demand/supply is assessable in terms of a cost that the system bears per its structure to carry out water demand/supply service administration. That system cost is a measure of system fit connected to the spatial externalities between human-tailored and natural governing systems (Moss and Newig, 2010).

To recognize the water management administration externalities (for demand/supply) in a basin, the research focused on constrained levels of administrative system, mapping them at spatial scale according to the arranged (administrative) service routes for water demand/supply. This scrutinizes if the existing jurisdictional boundaries and basin boundaries are reasonably admitting each other for the created water supply-demand chain flow. In consonance with this matter, **this study tries to assess the fit of the structure of governing system of water demand and supply associated to given administrative levels with relevant territorial/jurisdictional boundaries to a lake basin (hydrological system) through calculation of system cost**. This helps to evaluate if the levels of administrative scale are sufficiently able to internalize spatial externalities of water demand/supply issues and to lessen costs of the governing system. The outcome supports attainment of more fit mapping (Breton, 1965) of the governing system to the hydrological system.

Although the work mainly investigates the system issues due to the scales/levels interplay, the problem of scalar fit is also considered. Since, the flow of water (with relevant scalar organization) among jurisdictional boundaries associates with an institutional scalar organization (e.g. Pahl-Wostl et al., 2021; Moss and Newig, 2010), the fit of the system also deals with scalar dimension issues. Therefore, using methods from statistical mechanics, a branch of classical mechanics to investigate the behavior of systems, a specific tool taking into account both scopes is employed for the analysis.

Statistical Mechanics Methods (Gibbs, 1902) are considered as useful tools to study systems with uncertain knowledge about their status in socioeconomic and natural science domains. Adopting atoms collective behavior in a magnetic field, it is explainable how a collection of interdependent behaviors of entities in a (social) system determines the system status (Durlauf, 1999). Such methods are incorporated to create frameworks for studying interactions-based socioeconomic models (Durlauf, 1996). Even more, they have been used in the study of the population-wide characteristics in human-created systems encompassing the aspects of social networks (Arthur et al., 1997; Mantegna and Stanley, 1999; Contucci and Ghirlanda, 2007; Castellano et al., 2009; Kusmartsev, 2011; Barra et al., 2014; Contucci et al., 2017; Agliari et al., 2018; and Contucci and Vernia, 2020), methodological management (Braha and Bar-Yam, 2007), and political issues (Meyer and Brown, 1998) as well as decision making (Ortega and Braun, 2013; Bensoussan et al., 2013).

Since the problem of fit is mainly a system fit problem, the application of statistical mechanics models which consider scalar properties of motion signifying the system as a whole are remarkably beneficial. Relevantly in this work, the methodology braces the articulation of an analysis framework which portrays the effect of an adjusted configuration for water demand/supply interactions in given levels of an administrative scale via approximation by a single averaged effect able to lessen a many-body problem to a one-body problem. This helps conversion of a complex model to simpler one which its global behavior can be studied much easier (Seyedi, 2015). Focusing on an empirical case (a lake basin), we apply the methodology on a real-life setting in a particular region, furthermore, compare it with other assumable governing systems in that region. The research is set forth to detect the fit of the governing system for water demand and supply at certain provincial/regional levels to a lake basin.

## 2. Aim Conceptualization

To explore the problem of fit, selecting a lake basin as a case study area (please see next part), we concentrated on three levels of administration associated to spatial scales, namely: local, provincial and regional (Fig. 1). Positionally the basin is located inside three provinces each include relevant geographical zones (Fig. 2 and Fig. 3). The zones are administratively affiliated to the local level, and they are considerable as the entities demanding for water according to the needs of different users (here: Drinking, Agricultural and Industrial). In case of freedom in interactions for water demand and supply, different zones may interact based on their choices and decisions to make demands/supplies and therefore, the indeterminacy could accompany the related interactions. While this idea is majorly relevant to the efficacy of human choices at micro levels and individual users, at meso levels such as zones and provinces (concerning our case) the political-administrative arrangements try to control and orient the interactions into the ordered set-ups. This restrictive control desires to tend the interactions towards the 'determinacy' (almond and Genco, 1977). To implement a mechanism for such purpose, relevant to the case of this study, the governing system had incorporated particular entities responsible for responding the demands of the zones of distinct provinces where spatially located in the capital of each province. The aim was to vertically adjust focal entities institutionally armed with higher authority for dealing with demands/supplies. Administratively nailing those entities above zones, the governing system was largely able to constrain the interactions for water demand/supply to the channel between the zones with them and not with each other. Such entities which here we call them 'core(s)' were designed to cope with needs of disparate sectors. Thus, it is recognizable that each sector (e.g. drinking, agriculture) has its related core to organize the required supplies in that sector. Since cores institutionally/administratively set out above the level of zones, they are affiliated to the provincial level at administrative scale. Furthermore, the regional level encompasses more than one province with their relevant zones and cores. Beside arranging cores, another strategy the governing system get benefit of that for restrictive control and increase of determinacy in interactions is the political-spatial partitioning of the region into separate provinces with distinct jurisdictional boundaries in line with the organization of the overarching governing system of the country. This provision limits the interactions of entities into their affiliated province and embeds horizontal restrictive control on interactions. Conclusively, exploiting such strategies, the governing system attempted to set-up a controllable and more deterministic administration operation. Such determinacy may be achievable in some levels at temporal scale based on the political-administrative conditions.

Having those notions, to observe the spatial fit of water demand and supply governing system to the basin in a given period of time, the administrative flows of demand and supply were derived in the region. Correspondingly, adopting a statistical mechanics approach through applying a mean filed model (see parts 4 and 5), the study tried to map the associated interactions for water demand and supply systemized through administrative/institutional arrangements. Indeed, that mechanism is inserted through the forces which are devised by political/institutional settings. The interactions among cores and zones in this study were mapped according to the contemporary state at the time of analysis. The plans leading such interactions, nevertheless, are embedded at management scale which is linked to the national level of administrative scale. According to the constrained levels of administration in this study and spatial configuration of zones a cost function was formulated via mean filed models (parts 4 and 5) to measure the effectiveness of governing structure. Such cost function is a theoretical computational aid to measure the impact of the designed interactions, and forces for operating water demand-supply process. According to the extent of such cost the fit of the governing structure associated to relevant levels is identifiable at contributing scales.

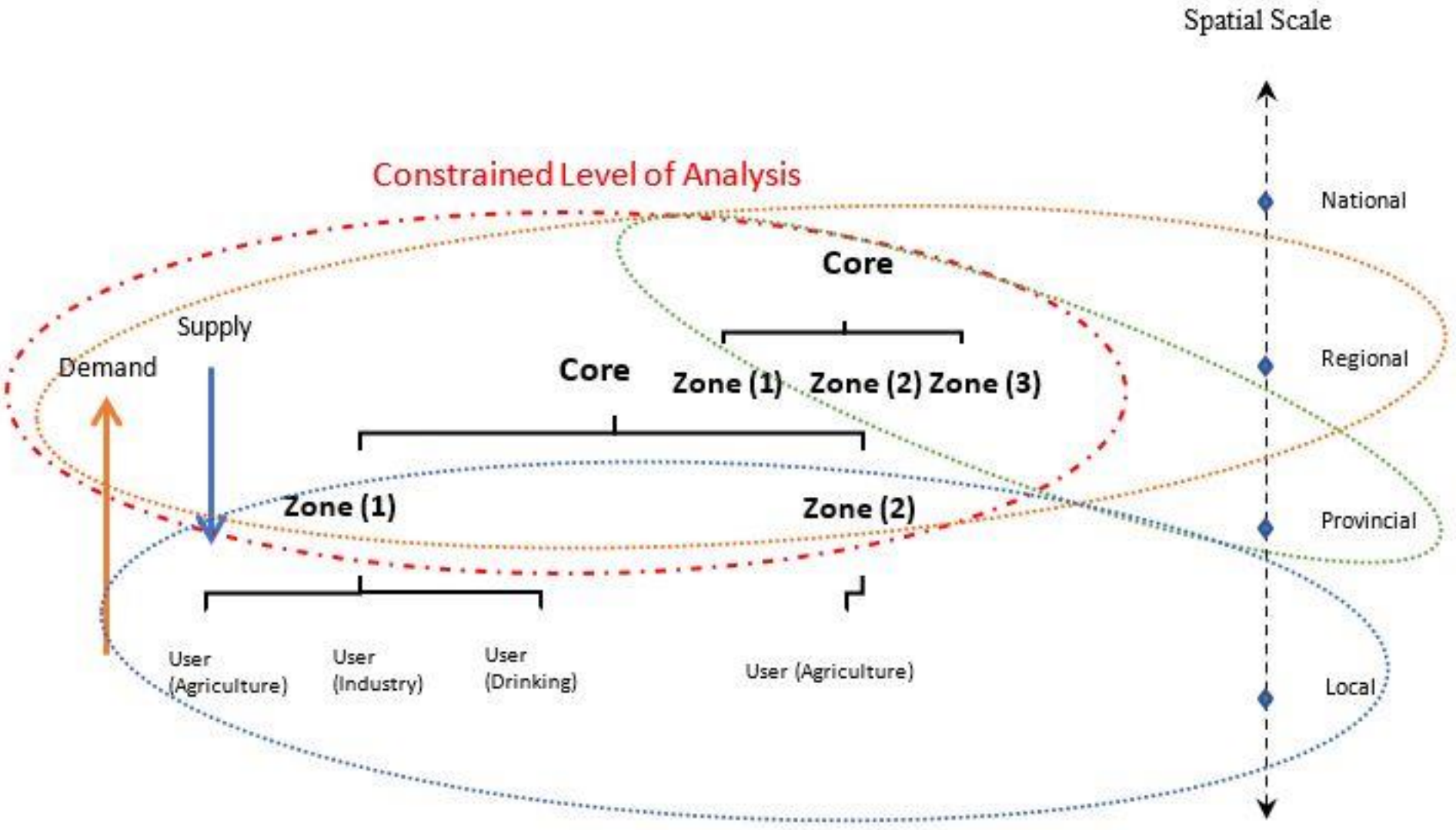

**Fig 1.** Schematic of the focused levels and scales relevant to the case study. A provincial level takes into account a core with its related zones and a regional level includes more than one core and their relevant zones.

## 3. Case study

To examine the model, a distinct region was selected in Iran. The Lake Urmia Basin with its 6.5 million inhabitants and arid to semi-arid climate is positioned among the six major basins in Iran. The rainfed agriculture and industry are the major economic sectors in the basin (Bakhshianlamouki et al., 2020). With the area of 52,000 km$^2$, and a hypersaline endorheic ambience, it is registered as part of the Ramsar Convention (1971) as well as a UNESCO protected biosphere in the year 1973 (Eimanifar and Mohebbi, 2007).

Due to the climate change; decline in precipitation and temperature increase, besides the sharp developments of barrage infrastructures such as dams and construction of a causeway across the lake (Eimanifar and Mohebbi, 2007), in addition to huge rise of ground water abstractions and other relevant issues arisen from inefficiencies of water conveying systems, and weaknesses in water management administration, the lake surface area has shrunk by over 80% during the last three decades (AghaKouchak et al., 2015). With 89% of available water use (Shadkam et al., 2020), the Agricultural sector is the largest consumer in the basin and respectively the Drinking and Industrial sectorsconfront are the next ones.

Whereas the basin is surrounded by two provinces named according to their geographical position to the lake: West Azerbaijan, located on the west side of the lake, and East Azerbaijan, located on the east side of the lake, it also includes another province, Kurdistan, which involves to the basin via a small territorial contribution while shares a substantial extent of water (Fig. 2). On the whole, there, water management organization is composed of the main ruler, Iran Ministry of Energy (MOE), which defines the policies and allocable water amounts from different sources, besides its provincial arms, the Regional Water Companies, which control and manage the water resources.

In addition, there are main distributers/suppliers such as Water and Wastewater companies responsible for drinking and sanitation aims; Agriculture organization responsible for irrigation and drainage initiatives; Industrial and Mines Organization which facilitates water supply for industrial zones; and department/organization of Environment takes care of pollution and ecosystem requirements. Those entities were following self-organizational objectives/targets until the year 2013 while started to take some shared responsibilities through the program of Urmia Lake Restoration. Generally, those organizations have their own representatives (local affairs) in main cities of the relevant province for water management, demand and supply initiatives.

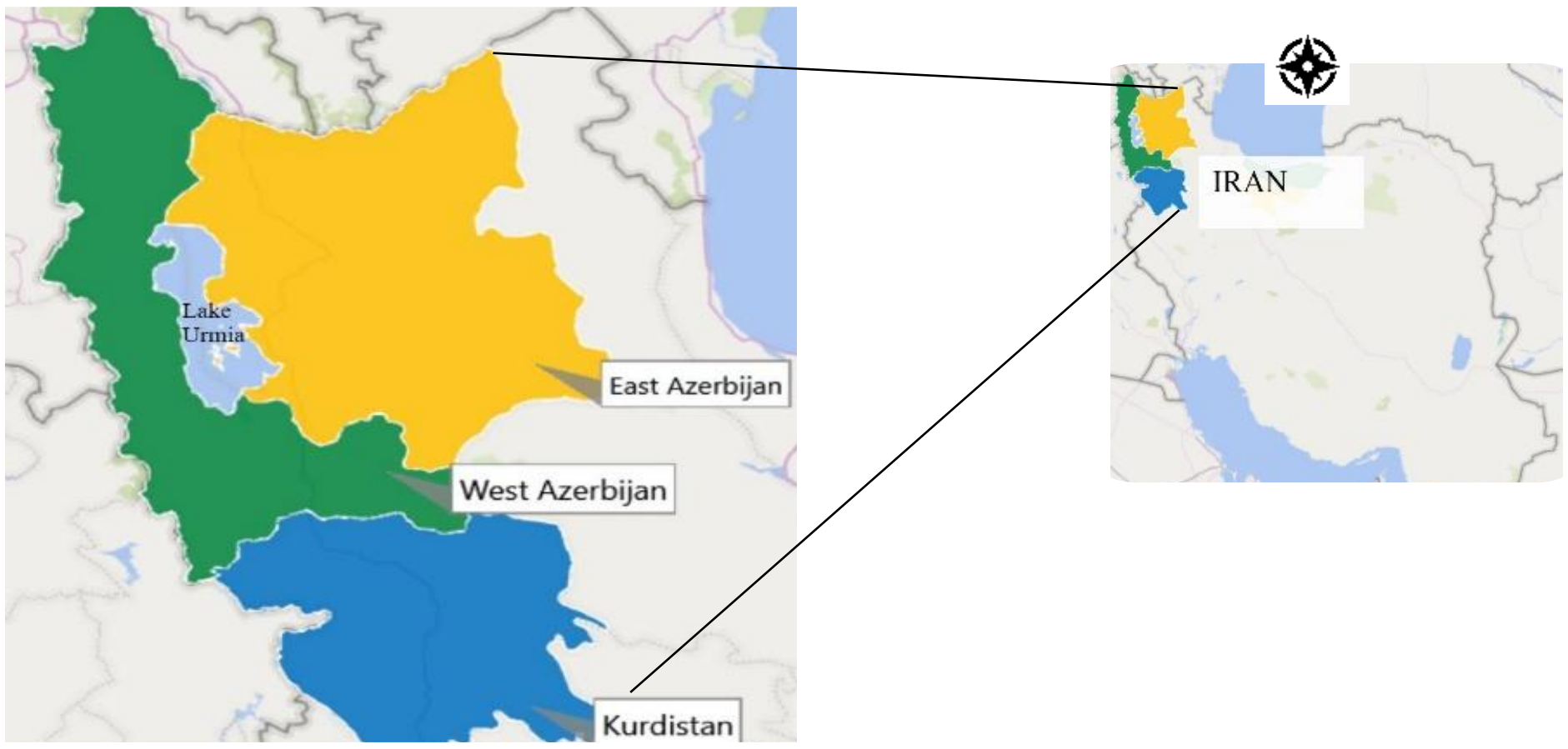

**Fig. 2.** The provinces surrounding the Lake Urmia Basin, Iran.

According to the available data (MOE, 2012; 2013a; 2013b; 2013c; 2014; and 2015), three main sectors are using water in the basin, and therefore we consider the entities of Agricultural, Drinking, and Industrial sectors in 25 geographic zones (Fig. 3) spread in 3 provinces[†] to tailor the analysis outlines. Each province has to manage the provincial issues inside its jurisdictional boundaries while may also communicate with another province for the transboundary issues through the highest administrative level of that province. This research is conducted based on the relevant data and existed administrative setting in the region until the year 2005, so, according to the official reference report (MOE, 2014) there hasn't been any water transmission from/into the basin until then.

---

[†] The two zones (numbers:10 and 25) were divided into two separate ones in terms of administrative belongings to the West and East Azerbaijan provinces, even more the capital of Kurdistan province (although it is not located in the basin) is administratively contributed to the base model (3-cores).

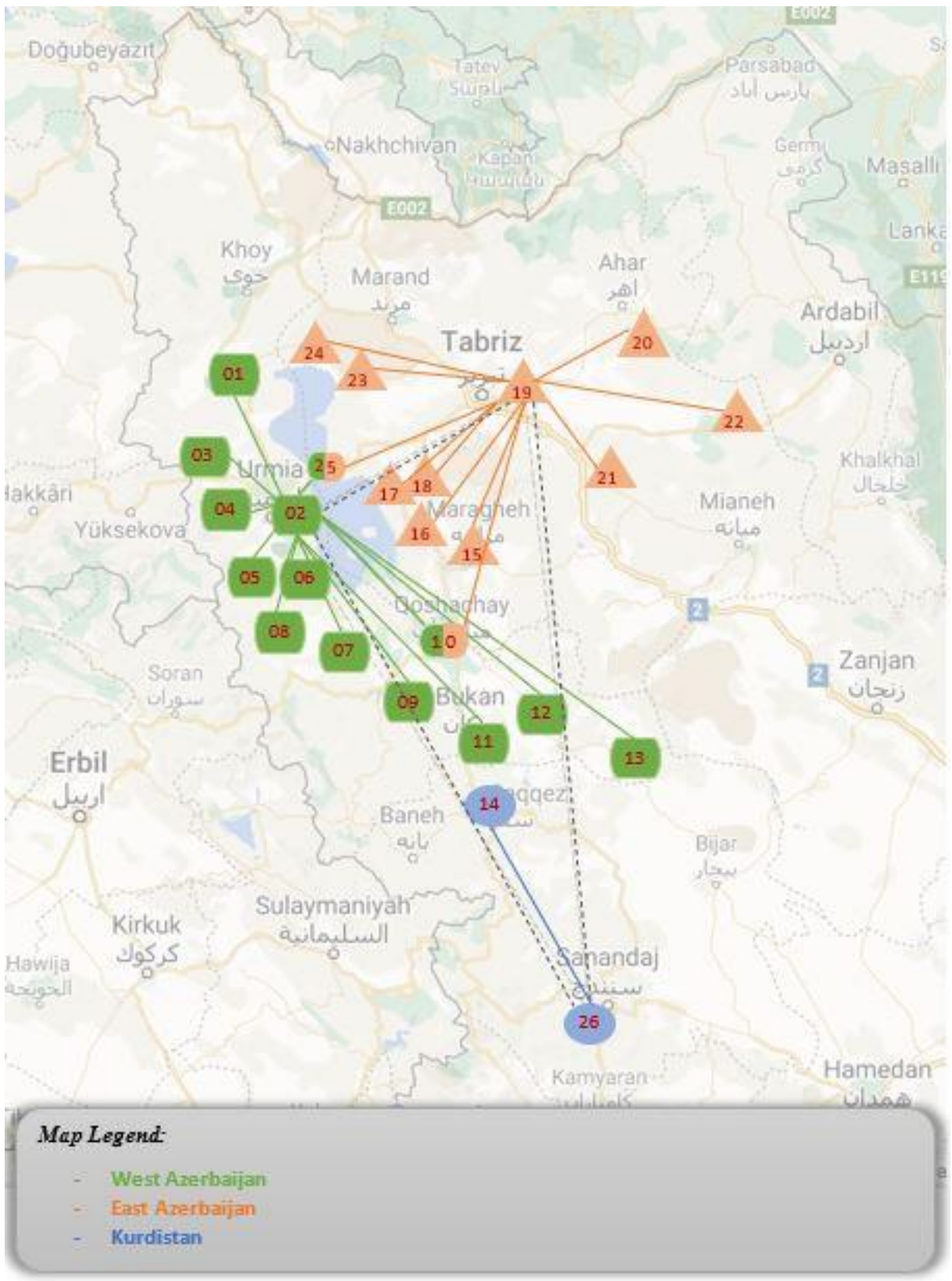

**Fig. 3.** The main zones of three provinces located in the Lake Urmia Basin.

## 4. Mean Field Models:

Within the scope of Statistical Mechanics, the Mean Field Models (MFMs) are among the largely exploited machines for studying socio-economic and political matters. MFMs apply explicit and exact computations so that the behavior of the complex systems can be studied easily. In other words, the method assists exploration of the macroscopic behavior of systems induced by microscopic interactions of their entities. Through the method it is possible to estimate the dynamics and connectivity distribution (patterns of interaction) of entities. The application of the concept in societal domains was originated based on the laws of thermodynamics capturing the behavior of atoms and molecules (Thompson, 1979 and Durlauf, 1996). In this study, we apply one of the successfully examined MFMs in studying the social and economic systems. The applied Curie-Weiss Hamiltonian model (parts 4.1 and 4.2) containing the high-occurrence-likelihood of interactions amongst some particles with opposite spin may play a prominent role as a suitable instrument for observing systems' semideterministic state very close to the boundaries of indeterminacy. Correspondingly, through Hamiltonian formulation of the governing system of water demand/supply, we aim to evaluate the extent of system's fit at the basin scale.

*4.1. The Curie-Weiss model*

The idea of mitigating a multipartite system to a one-population problem (Mean-Field Theory) by replacement of all interactions to any one population with an average/effective interaction (Molecular Field) was appeared initially in the work of Pierre Curie and Pierre Weiss when they tried to explain a simple classical system that exhibits phase transition (Curie, 1895; Weiss, 1907). Among several MFMs, the Curie-Weiss Approximation is one of the frequently applied models in studying social, economic and political issues (e.g. Barra et al., 2014; Contucci et al., 2017) in which an interacting system of $N$ particles (entities) is generally Hamiltonized with uniformly binary random spins ($\sigma_i$) as:

$$H_N(\sigma) = -\frac{J}{2N}\sum_{i,j=1}^{N}\sigma_i\sigma_j - h\sum_{i=1}^{N}\sigma_i, \qquad (1)$$

where the magnetization of the binary configuration, is defined by $m_N(\sigma) = \frac{1}{N}\sum_{i=1}^{N}\sigma_i$ along with the value of the external force: $h$, as well as the interacting positive constant: $J$. This means that the interested multispecies Hamiltonian of the Curie-Weiss machine may be characterized by

$$H_N(\sigma) = -N\left(\frac{1}{2}\sum_{i,j=1}^{N}\sum_{i,j=1}^{N}\alpha_i\alpha_j J_{ij} m_i m_j + \sum_{i=1}^{N}\alpha_i h_i m_i\right) \qquad (2)$$

with assumptions of the relevant magnetizations:

$$m_k(\sigma) = \frac{1}{N_k}\sum_{i=N_{k-1}+1}^{N_k}\sigma_i\,;\ k = 1,\dots,\mathcal{T};\, i = 1,\dots,N;\ N_0 = 0 \qquad (3)$$

along with their proportional subsets' size, $\alpha_k = \frac{N_k}{N}$ ($k$ is the number of sets) where the first binary choices selected from $\{\sigma_i |\ i = 1,\dots,N_1\}$ when the last alternatives are: $\{\sigma_j |\ j = N_{k-1}+1,\dots,N_k; k = 2,\dots,\mathcal{T};\ \mathcal{T}\ is\ the\ total\ number\ of\ partitions\}$ with the fixed term $N_1 + \cdots + N_k = N$. Such that, the relevant vector field $h_i$ is assigned by the $k$ magnetic fields: $\begin{bmatrix} h_1 \\ \vdots \\ h_k \end{bmatrix}$, while $J_{ij}$ tunes the mutual interaction among those particles with respect to the following adjusted matrix:

$$\begin{bmatrix} J_{11} & \cdots & J_{1k} \\ \vdots & \ddots & \vdots \\ J_{k1} & \cdots & J_{kk} \end{bmatrix}$$

in which the positive blocks $J_{ij}$ reveal the interaction among the particles belonging to the same/different subsets (Contucci et al., 2008; Fedele et al., 2013). In other words, the concept communicates that through maximization of the pressure, the output of self-consistent equation of a multi-species system pictures the total energy exchange in the system. Furthermore, it transmits that the relevant magnetic critical/threshold exponents are discernible as a function of an effective mean-field (Contucci and Ghirlanda, 2007).

*4.2 Meanings of parameters in the case*

In a set (province), to see for a moment the behavior of an semideterministic system which its determinacy is provisioned through political-administrative arrangements (parts 2 and 6), we consider some almost surely interaction events $J_{ij}$ (holds amount of water demand/supply) between entities (e.g. zone $i$ and core $j$, and vice versa) with opposite paired 'spins' or administratively we call 'orientations', $(\sigma_i, \sigma_j)$, which may occur under the influence of the diffused institutional/administrative rules of interaction (part 6). Such orientations resemble the arranged 'trait values' for the entities in a demand/supply chain process based on the discussed instructions.

Within tending realms towards determinacy, in a system composed of subsystems/sets, according to the network analog of Gauss's law that relates a measure of flux through a set's boundary to the connectivity among the set's nodes (Sinha et al., 2018), each entity is surrounded with a magnetization (connectivity flux) rate $(m)$ relevant to its belonging set. Through this notion and in line with the principals of MFM, the magnetization in a set is achievable by averaging all entities' signs in that set: $m_\sigma = \frac{1}{N}\sum_i \sigma_i$. This let the reflection of that magnetization, '$m_\sigma$', as 'a group (provincial) trait value' per a certain set which can be allotted to all entities of that set characterizing the set as a community with appropriate flux through its boundary. Correspondingly, the proportionate 'subset sizes', '$\alpha_k$', are calculable dividing the number of entities in a set (province) over the number of all entities in the system (basin). Like as the magnetization each subset size is equally allottable to all entities of the relevant set.

As a result, the total cost function in multi-populations can be expressed as Eq.2 when the relevant mutual interactions and the associated provincial trait values $(m_\sigma)$ are expected by the collection of possible configurations (jurisdictional/administrative in our case) in a system. As conceivable, provincial trait values and their relevant subset sizes in the basin are linked to the governing system structure. They stand beside the extent of demand/supply (can be seen in Eq. 2) to formulate the interaction rates of entities in a certain set. Therefore, those variables are nominated as control variables could be used in appropriate structuring of a system to improve the match between administrative and biophysical systems. However, they also reveal that how a system could be exposed to influences by political-administrative provisions. Apart from the discussed parameters and variables contributing to Hamiltonian formulation of the system, we go through with the external forces $(h)$ which have hidden (implicit) attributes. Those forces result from the combination of rules formatting the context in which the interactions for water demand and supply take place. In the environment we assess the interactions, such forces are noticed as effects associated with the set of instructions structuring the administration mechanism. Therefore, they can be intuited as a sort of institutional forces trying to sustain the system dynamism while orienting the interactions. Such forces are perceptible as pre-existing ones if we refer to them at the time of analysis. However, they may change in time according to the changes in the system composition by the decisions in pertinent levels of administration. The abstract of those forces could be well quantified by inverting the abovementioned many-body approximation (Eq.2), (for further information please see Fedele et al., 2013):

$$h_i = \tanh^{-1}(m_i) - \sum_{j=1}^{N} \alpha_j (J_{ij}) m_j \qquad (4)$$

The Hamiltonian cost $(H)$ is a cost a system bears per its structure. It resembles the total cost a system carries for administration of all interactions in the system. In fact, the costs related to a given entity in a particular set are symbols of state parameters that show how a system and/or an entity is affected by the existed structure. Another parameter beside the cost which supports the estimation of the state of a system, is the Pressure $(P)$ which is determined by the logarithmic format of the exponential formulation of the symmetric position for this Hamiltonian (for further information please see Seyedi, 2015):

$$P_N = \frac{1}{N}\log(\sum_\sigma exp(-H_N(\sigma)) \qquad (5)$$

This parameter is tightly connected to the cost. It reflects the measure of "return" a system receives per the cost it bears. Whether a system can better preserve its energy via lessening its cost, its pressure or return will grow higher. $P$ for the system is discernible as a type of return, interlinked to system's (dis)organization. Whereas, for the entities of the system, it denotes a sort of gain affected by the cost intertwined with the system structure. In case this gain is reasonable enough or good, it can be sensed as a profit or benefit for the zone. To transmit this apprehension, the values of $P$ were projected in radar charts resembling the gain of a zone as an inflation. As later will be discussed further in this paper, the parameters $H$ and P, are the outputs of the designed machine and the analyst diagnosis the system status per different structures via such state parameters. Having those notions, the model development is further discussed in the next part.

## 5. Tailoring concept to the case

As indicated in part 2, the zones are considered as demand entities in local level which themselves include the water requirements of three sectors according to the MOE report (2014). Each sector (Agricultural, Industrial and Drinking) requests for its demand from the responsible entity (core) in that sector (part 2). The core is administratively above the level of zone. Cores resembling the provincial level entities that can surely interact with the zones of their province and may interact with each other for information exchange. However, according to the administrative restraints the zones almost never can directly interact with each other for their requirements. Hence, the structure of water demand/supply governing system for a given sector in the year 2005 at Urmia Lake Basin concluded in three cores associated to three provinces responding the demands of their zones (Fig. 4a). Furthermore, due to the distinct scalar dimensions of the water demand/supply volumes of each sector and the sectors' administrative disconnection of water demand/supply process, the related interactions of one sector are assumed to be valid for that sector only, and accordingly the interaction rates of disparate sectors should not be compared with each other. Beside the existed assembly of governing structure (3-cores) defined based on jurisdictional boundaries of the three provinces, we evaluated the two other forms (Fig. 4b and 4c). Those forms represent the state in which all zones may interact with one hub for their demands (1-core) as well as the state that the only contributing zone of Kurdistan province (the zone 14) could be included in West Azerbaijan's team and the basin administration could be shared between the two, west and east Azerbaijan provinces (2-cores). Placing these postulations in addition to the practical experience of the main author in the region[‡] , we analyzed the system per the three constitutions through the application of the Curie-Weiss Model.

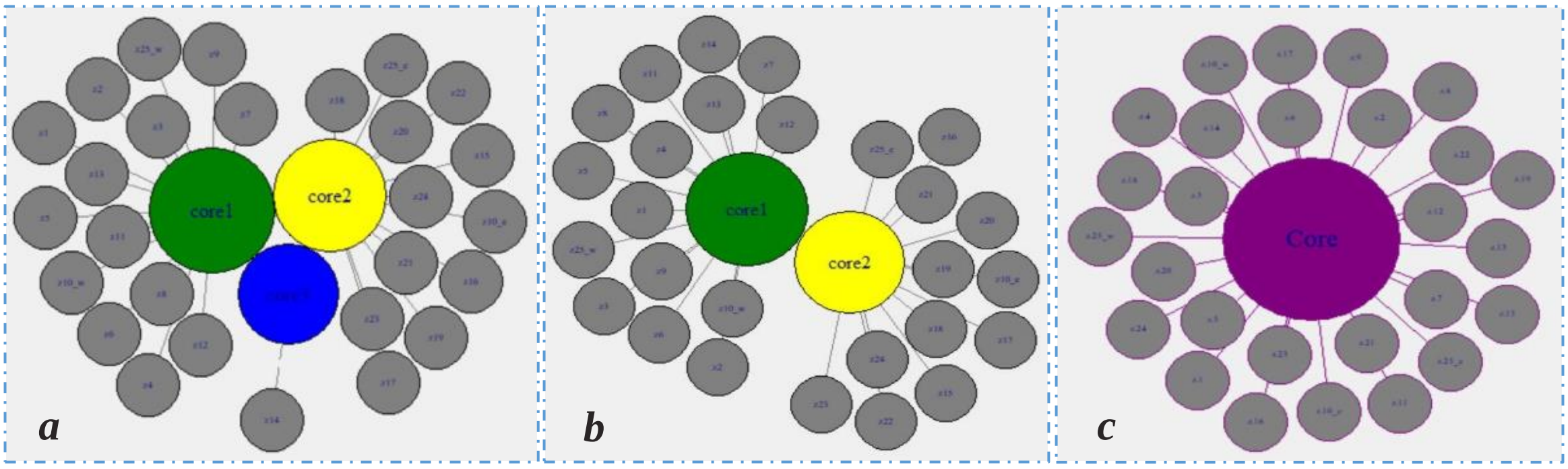

**Fig. 4.** The entities partitioning in 3 modeled system forms: (a) 3-cores (left), (b) 2-cores (middle) and (c) 1-core (right).

‡ Technical Expert at West Azerbaijan Province Water and Wastewater Company (Shirzad et al., 2011)

## 6. Modeling the governing system

According to Durlauf (1996), the collective interdependent behavior of entities in a socio-economic system can be adopted from collective behavior of atoms in a magnetic field. Furthermore, Sinha et al. (2018) proved that the Gauss's law (Gauss's flux theorem) is also applicable for social systems and the connectivity flux in a social network is the implication of electric flux in a magnetic field. Therefore, the concept of electric charge was let to determine the probabilistic connectivity distribution among contributing entities according to the administrative/institutional connectivity rules: i) zones can almost never interact with each other; ii) in a certain province each zone can almost always interact with its relevant core; iii) cores may contact each other for information; iv) sectors should be treated separately. Correspondingly, each zone of distinct province was most likely connected to its core appointing an opposite paired binary connectivity (charge) sign, $(\sigma_i, \sigma_j)$ to them (the ratified trait values which let the demand/supply flow be defined according to the routes instructed by administrative provisions under given political-administrative conditions).

Administratively characterized in our example case, the zones interact for water demand with the extent of demand $J_{ij}$[§]. Such extent is the volume of water solicited by a zone which is rescaled in our case separately for distinct sectors respective to their max-min demand amounts based on Feature Scaling method to place pertinent and rational values incorporable in the evaluation device (Fig. 5). Therefore, the rescaled demand/supply amounts of a particular sector $(J)$ are only comparable within that specific sector and not among the sectors. Similarly, the cores interact to respond the zones via the extent of supplies ( $J_{ji}$) which are rescaled based on the same method.

In obedience to that inception, at the first stage, we let the interactions be shaped through a predefined concept of magnetization based on adaptation of Gauss's law for networks to the existed administrative structure in the basin. According to the governing structure, the contributing entities partitioned in three sets respective to the present provincial boundaries in the basin. Each set includes a core, besides its zones forming the total number of contributing entities positioned in that set $(N_{\text{i}})$. The total number of entities $(N)$ contributing to the basin's water demand and supply initiatives is considered as the sum of all entities in the sets participating in interaction process $(N_1 + N_2 + N_3 = N)$. Mapping the entities' configuration, each entity was featured by a trait value $(\sigma_i)$ based on the rational of attraction (connection) due to opposite signs (trait values) between the core and its zones to place the interaction scheme at the time of analysis. However, in relation to the Kurdistan province since its core spatially does not belong to the basin area, it is assumed as an electric pole with no particular sign (neutral) supporting the administrative linkage of its only zone in the basin to the system. Above all, the overall order of the signs (orientations/trait values) of all contributing entities endorsed the connectivity in the system. Subsequently, the provincial trait values $(m_\sigma)$ as well as their proportional subset sizes $(\alpha_k = \frac{N_k}{N}, k = 1,2,3)$ were calculated for each set (province). Tuning the 'interaction rates' $(\alpha J m)$ via the demand and supply (rescaled) volumes $(J_{ij})$[**], respective to the objectives of the entities in our example case, the external forces were determined according to the Eq.4. Furthermore, portraying the administrative interaction episodes in presence of interaction rates ($\alpha Jm$), external forces ($h$), and the participant zones of each set (Fig. 7), the effectiveness of the governing structure for water demand and supply interactions was noticed.

---

[§] $J_{ij}$ is a number that summarizes the nature of the interaction between i and j, for instance, if i is a solicitor entity, it interacts with the extent of its demand as the value of $J_{ij}$ for its interaction.

[**] The volumes of demand and supply (MOE, 2014) are rescaled for each sector separately respective to the distinct sector's max-min amounts based on Feature Scaling method to place pertinent and rational values incorporable in the evaluation device. Therefore, a particular sector's rates are only comparable within that specific sector.

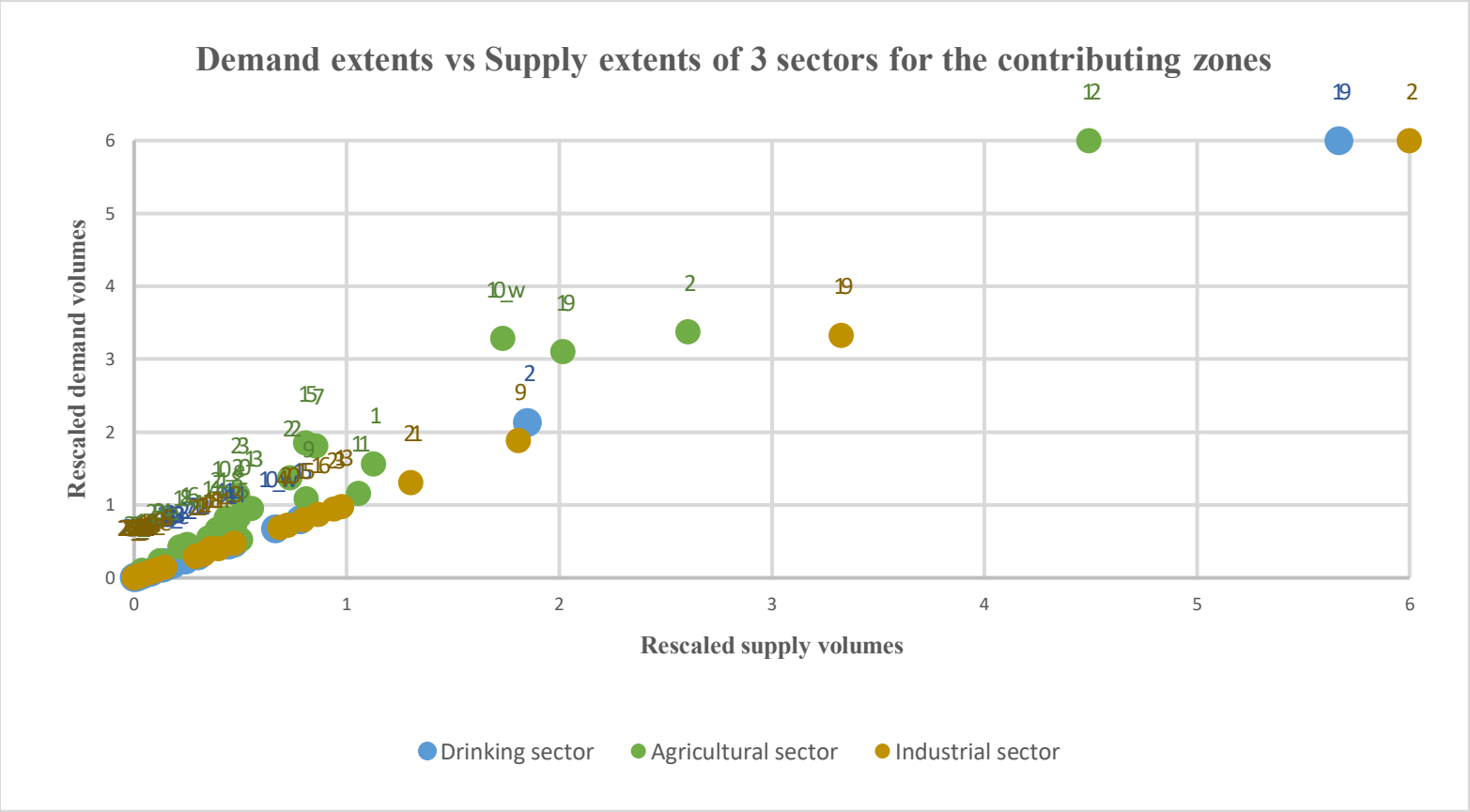

**Fig. 5.** The rescaled rates of water demand and supply volumes (MCM) of three different sectors for the zones of the Lake Urmia Basin (MOE, 2014). Those rates manifest the biophysical scalar organization in an administrative mutual interaction plane (for demand and supply). The zones are pinpointed through their relevant numbers (above dots) for different sectors.

In the second stage, the relevant costs ($H$) of each sector per different system structures were determined through the Eq.2 (Fig. 8). Lastly, in the third stage, the system and zones pressures ($P$) were measured via the Eq.5. As mentioned, the pressure for zones were discerned as the quintessence of gain (Fig. 9).

According to the induction in part 5, the model basically was designed for the 3-cores structure respective to the contemporary governing system in the basin. However, to analogize the existed system with some other relevant structures, the model was customized for the two hypothetical forms, namely the 1-core, and 2-cores. In the 1-core setting, the jurisdictional/provincial boundaries are neglected, and the zones are interacting with only one core (for each sector). While, in the 2-cores setting, the only located zone of Kurdistan province (14) in the basin is assumed to be an entity of West Azerbaijan province.

## 7. Results and discussion

To get the picture of interactions, effects, and outcomes per different system settings in our case, we split the analysis in three clusters. Duly, the first cluster goes through with the type of control elements in rule-structured environments which are created through instructions for administrative interactions in the Basin. In this cluster, the ensemble of administrative interaction episodes is projected (Fig. 7). The aim of this projection is to show how system composition launches an area corresponded to the interactions and forces in which the entities of a certain set are configured according to their rates of interaction alongside the forces (Fig. 6). The more sited interactions inside their relevant area, the effective the system structure, since it has handled more interactions. Indeed, the interaction of each zone of a set which is not sited in such an affiliated area, is not handled by the designed structure of the system. Therefore, the analyst can detect how the control or handling capacity of a system (to deal with its zones' interaction rates) varies by its structure. Displaying the results of this cluster in Fig. 7, we narrate the outputs correspondingly.

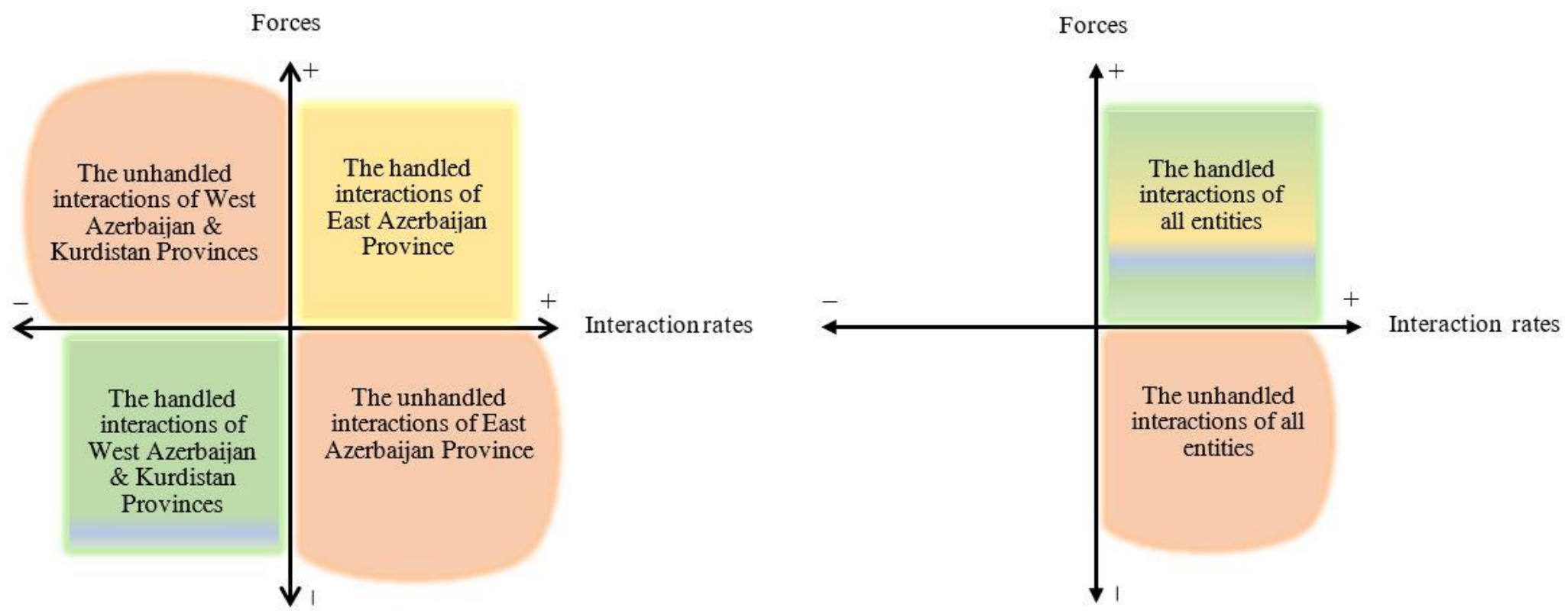

**Fig. 6.**The launched areas by the system composition for interactions of different sets per the 3 formations; right: the 1-core and left: the 3-cores (also 2-cores).

To give a clear-cut induction on the results of the first analysis cluster, we emphasize on an affiliated area to the defined set of entities in different configurations. According to this knowledge, in the 1-core system, there is only one associated area (Fig. 6) to all interacting zones in which both interaction rates and external forces have the same sign (positive). Whenever a zone's interaction rate surpasses a critical value, it outstrips the appurtenant area and stands beyond there. This means that the proportionate handling capacity of that system determined by its structure is not effective for administering such zone's interaction rate. This issue is well detectable regarding the interactions perched on the below part of the horizontal axis of the 1-core setting. Another principal is that the values of forces in such cases have the reverse alignment with the zones' trait values ($\sigma$) meaning that those forces are struggling to orient such interactions toward their pertinent sphere, and this causes the force direction to be altered to the opposite one which is associated with an additional cost infliction to the system. Identically, in the 2-cores and 3-cores settings, the handled interactions of entities are configured in top-right and bottom-left portions where both interaction rates and external forces have the same sign. They are the pertinent launched areas for interactions in those systems. Like as the 1-core's state if an interaction placed in an area that is not affiliated to, it represents a sort of problem which arises due to the system's structural fit in handling the zone's interaction rate in that configuration.

Through this illustration and adopting a sectoral approach for the comparisons, we assessed the outputs of this cluster. The results (Fig. 7) indicate that once there is a 1-core system responding to the demands of all entities, the system is apparently impotent in handling the interactions of several zones, particularly in Agricultural sector. More indicatively, the interactions of zones: 1, 2, 7, 10_w, 12, 15, 19 are perching beyond their assigned area. Whereas, only, one interaction in 2-cores, and two interactions in 3-cores settings are unhandled. Apart from Agricultural sector, indeed the unfit of the 1-core system is realizable in other sectors. While the judgment comes to the 2- & 3-cores settings, the competition is tight. It is transparent that not only regarding the number of the interactions settled in the belonging areas, the 2-cores setting shows better fit, but this structure also handles its zones' interactions with more satisfactory rates. For instance, to handle the interaction of zone 12, in the 2-cores set up, the related force is smaller than the 3-cores set up. In addition, the interaction of zone 10_w is hardly being handled in the 3-cores setting but is well handled in 2-cores one. Correspondingly, a message that may stem from comparisons is that the predefined (political/jurisdictional) boundaries used for administration of water demand/supply (in 3-cores system) may not fit as the 2-cores system in the basin. This shall propose a slight reconfiguration/rescaling (Moss and Newig, 2010) and adaptation in administrative structure to benefit the system.

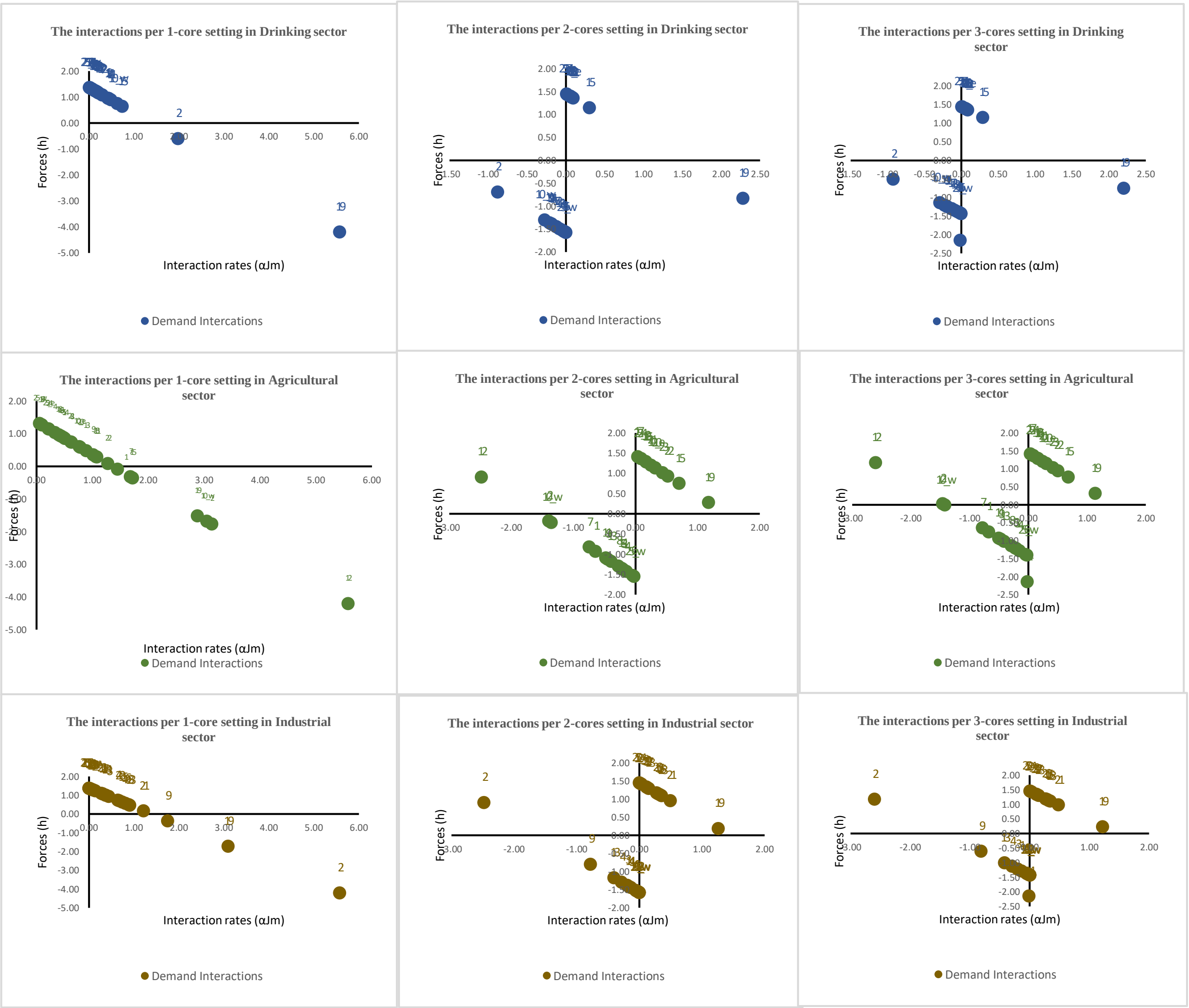

**Fig. 7.** The interaction episodes, interaction rates, and forces. Although we modeled the example case based on demand and supply interactions, to prevent confusions, we only displayed the attributes of demand interactions in the graphs. Even more, the demand rates are greater than the supply ones (in our case) which notify that if a demand interaction of a zone is settled in the affiliated areas its relevant supply interaction is already settled there due to its smaller rate. Each dot shows the interaction of a zone which is displayed by its number (above dot).

To further endorse the disclosures of the first analysis cluster, we deploy the results of the second cluster which compares the extents of costs per different system structures (Eq.2). According to the computed values, the system with 2-cores in all sectors dominantly takes care of its costs. In other words, the system has the smallest cost rates in different sectors for administration of the interactions comparing to the others (Fig. 8). As a matter of fact, the Hamiltonian could be a benchmark for the system fit. It communicates that the two compartments of Eq. 2. shall be harmonized in a way enlarging the rate of the function as a whole to lessen the cost due to the invert sign in the equation. The higher the rate the lesser the cost, the better the system fit. To achieve this aim, the system structure shall be formulated in a way that the interactions could be handled properly. Verily, the results of this cluster further revitalized the outputs of the first cluster by admiring the conduct of 2-cores system.

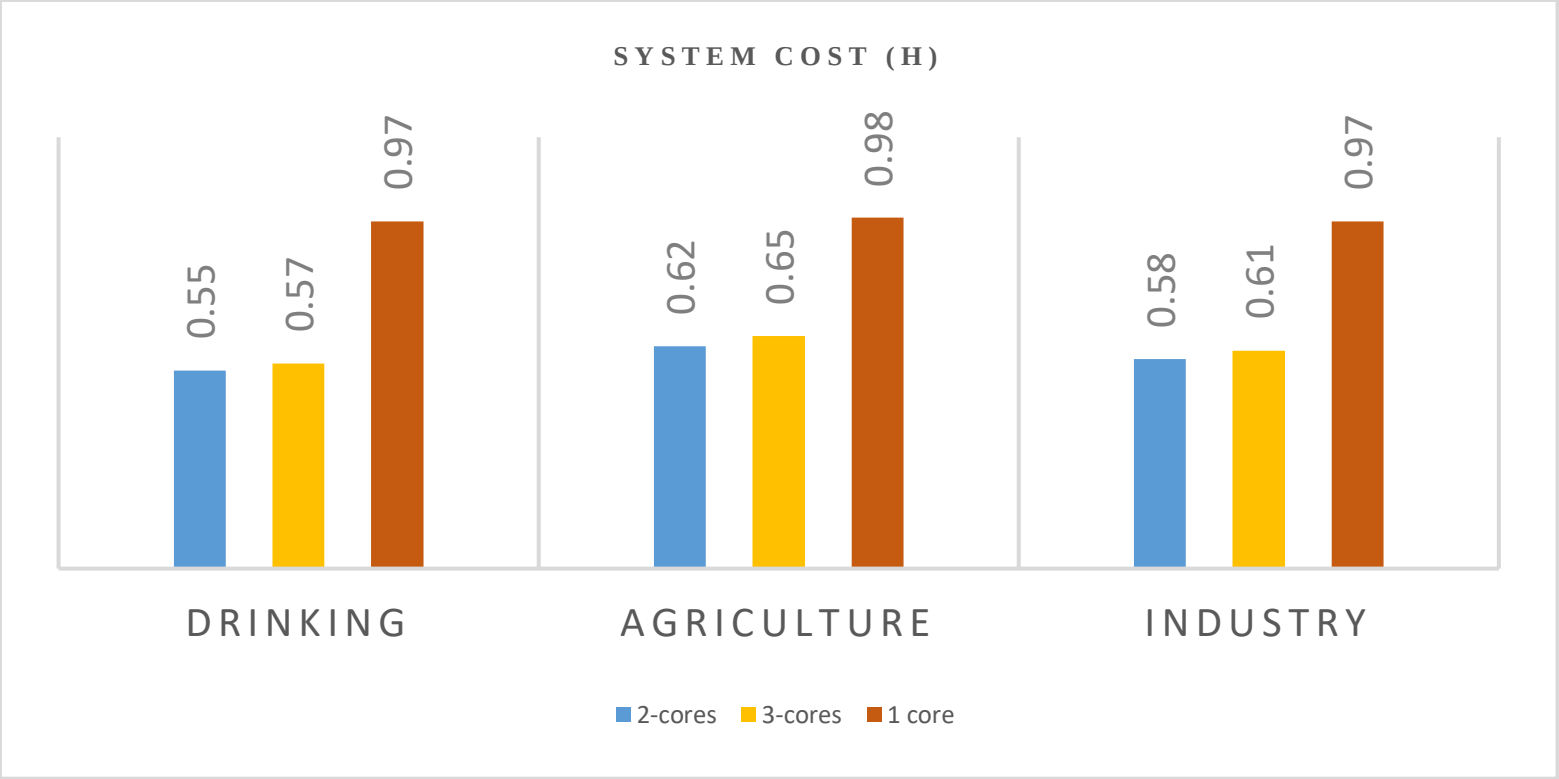

**Fig. 8.** The normalized rates of *(H)s.*

Finally, the third cluster of analysis tackled with the consequence of various structures and their costs on the state of the systems and their entities. The outputs of the third cluster (Fig. 9) revealed that how the zones gain valuation in different system settings. Such values resemble 'gains' from system's composition. In this cluster, although the pressure (return) of the systems didn't change greatly, specifically for some sectors (e.g. Agriculture), but the entities pressure (gain) vary significantly in different settings. Indeed, the system pressure per disparate sectors in distinct structures was resulted as a function of summary measure of cost values to all the entities in the system. In other words, it is a cumulative score of the range of all costs stems of system composition. Hence, in 1-core set for instance, the core held the large extent of pressure and the zones become less gained ones. Whereas, in 2-cores and 3-cores the gain is moderately diffused to the zones. This denouement rendered that in 1-core set it is the core's gain that mainly determines the system pressure as it possesses the mastery role, and zones are experiencing little gain, while in 2-cores and 3-cores settings this effect is unlikely. In detail, recognizing the results, the values of 2-cores system disseminated that again this formation deploys satisfactory gain rates comparing to the others.

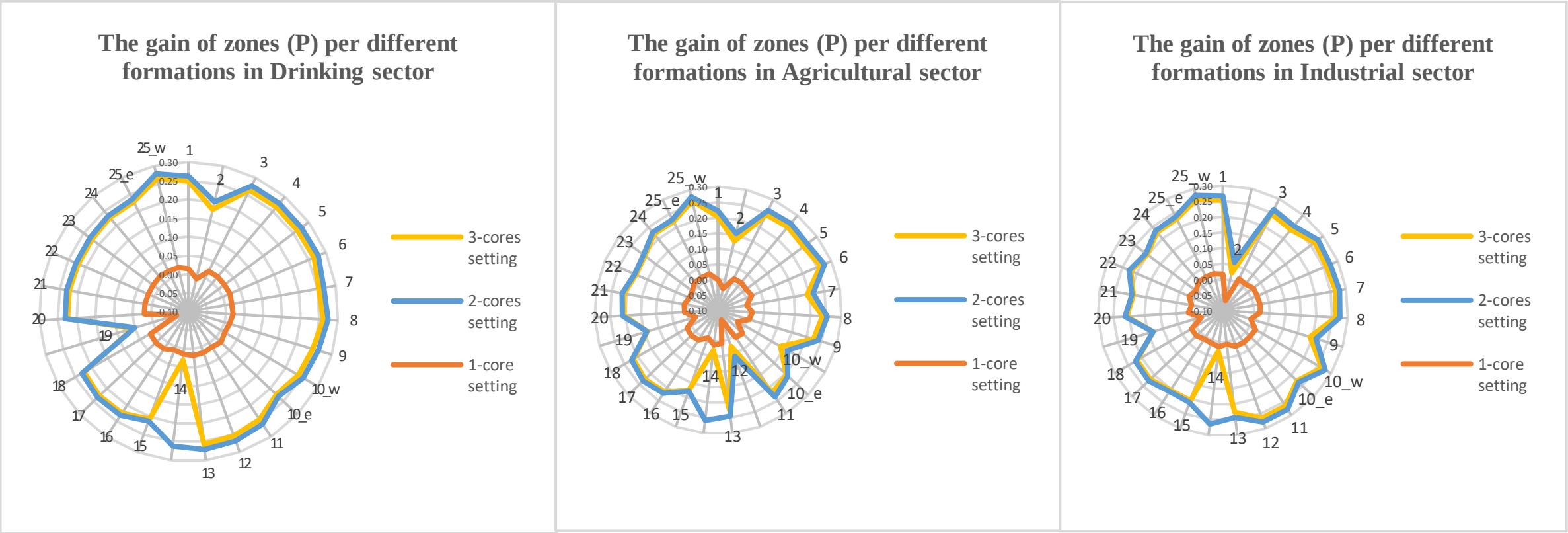

**Fig. 9.** Normalized rates of the pressure of zones *(P) per different* formations in three sectors

Transparently, apart from the apparent supremacy of the 2-cores setting to the 1-core, remarkably, in the zone 14, there is a huge revamp in gain comparing to the 3-cores system. This confirms that in spite of the existed structure (3-cores) in the basin, the zone 14 could be benefited if it was coupled with West Azerbaijan's group. This issue may further support boundary revisions usefulness in the basin.

Other than the aspects discussed in different clusters of analysis, another point worth mentioning, is the importance of the extent of demand ($J$) in different sectors which may lead to improvements in policies and decisions. In this regard, it is clear that the Industrial and Drinking sectors are satisfied predominantly while Agricultural sector strives to reach by its supply targets. Although, Drinking and Industrial sectors got higher priority in receiving their demands, it is the Agricultural sector that determines the overall status of the system due to the greater amount of water demanded in this sector. Therefore, a particular attention on this sector is necessary.

Another substantial fact is that the interactions of some zones cannot be handled through any of the structures. For instance, in Drinking sector the interaction of zone 19, or in Agricultural sector the interaction of zone 12 are not handled through the featured settings. Else, in Industrial sector the status of zone 2, not only exposes the same characteristics of the two beforementioned zones but it also subjects to delicate situations in Agricultural and slightly inferior in Drinking sectors. This may trigger an idea that there might be no best solution of partitioning to form the systems (horizontal arrangements) but there could be several good ones. Nevertheless, the vertical modifications in administration system may boost the fit of the governing system which their assessment overwhelms the content of this paper, and another extensive research is necessary to investigate their impacts.

## 8. The Outcome and Conclusion

Extending a pragmatic and analytical approach, three different systems for administration of water demand/supply interactions were investigated through the unified language of mathematics based on a physical principle. For an inspiring evaluation, the methodology is applied to an example case tuned by related political-administrative arrangements to restrict the indeterministic interactions. Exhaustively, through the probabilistic Hamiltonian mechanics' formation of a multipartite system, a water interactive-based societal case is prescriptively analyzed in order to assess the fit of a governing system for water demand/supply to the relevant basin. The research aimed to avail a unique window for the observers to follow the system fit per the interaction rates for water demand/supply along with the generated forces due to different circumstances in the region during the given period of time. In fact, the model enabled us to put together the semideterministic interactions per units of water resource in administrative-political settings with the product of exogenous factors in the form of forces under the relevant conditions. Furthermore, it supported the comparison of the existed system with two other imaginable ones. This uncovered that how the modus operandi of those interactions is affected by the system structure within distinct sectors.

Through the comparison of the three studied systems, it is revealed that even though the 3-cores system (contemporary) is moderately fit to the basin, it is not as proper as the 2-cores system in respect to the number of privileged zones and system cost. Even more, in 2-cores setting the interactions are better handled. Indeed, the 1-core set showed unmatched characteristics in all the sectors. This outcome highlighted that either the political/jurisdictional boundaries (Nabavi et al., 2017) or basin approach may be revised for fitter water demand/supply administration systems in the basin. On the other hand, the study shows that whether, the governing structures are deformable entirely or partially, the system can be enhanced accordingly. Therefore, during restructurings, it is recommendable to evaluate the configurations producing less costs as well as more benefits for the zones.

Results of the study may also divulge some additional intimations regarding systems' fit. Those revealed that in a system without perfectly fit structure for governance, the gain (pressure of the system) settles only in the cores but not reasonably in other components (e.g. 1-core setting). This issue would lead to malfunction of the system and/or deficiency while if the gain could be rationally diffused to other components, system hits rewarding results.

A notification may appear here, is that the number of cores has to be determined according to the system requirements and more cores doesn't exactly mean more fit. Of course, structuring a system for a particular objective such as water demand/supply will be more admirable if both vertical administrative arrangements and horizontal spatial provisions could be balanced jointly. Even more, the perfect fit shall respect to the objectives of the other systems (e.g. energy and food) in which the same entities may interact. Therefore, before any action in the imposition of a new system, the mutable and affecting circumstances shall be assessed rigorously embodying a sophisticated approach to restrain the eventual affections.

Openly, this paper discussed an example for a simple application of MFMs on a political-administrative structure of water demand/supply governing system during the year 2005 in the Lake Urmia Basin. This paper tried to show how such machine can assist exploration of more match structure for administration of the water demand/supply in the basin. Indeed, a comprehensive study may be necessary combining diverse objectives deemed for a given territory to appropriately formulate the levels on administrative scale, beside the hydrological one.

## Acknowledgements

We thank Alberto Montanari, and Gül Özerol for their constructive comments on the very first version of this manuscript as well as two anonymous reviewers whose comments helped to enhance the clarity of that. The authors also wish to express their appreciation to the "University of Bologna", and "Iran's National Elites Foundation" for funding the research.